\documentclass{sig-alternate-2013}

\usepackage{graphicx}
\usepackage{url}
\usepackage{mathptmx}
\usepackage{amssymb}
\usepackage{amsmath}
\usepackage{subfigure}
\usepackage{hyphenat}
\usepackage{color}
\usepackage{textcomp}
\usepackage{enumitem}

\usepackage[show]{chato-notes} 

\usepackage{cmm-greek}

\newcommand{\hide}[1]{}
\newcommand{\xhdr}[1]{\vspace{1.7mm}\noindent{{\bf #1.}}}

\newcommand{\cpt}[1]{\textsc{\MakeLowercase{#1}}}

\newcommand{\ie}{\textit{i.e.}}
\newcommand{\eg}{\textit{e.g.}}
\newcommand{\cf}{\textit{cf.}}
\newcommand{\etal}{\textit{et al.}}

\newcommand{\Secref}[1]{Sec.~\ref{#1}}
\newcommand{\secref}[1]{Sec.~\ref{#1}}

\newcommand{\Eqnref}[1]{Eq.~\ref{#1}}

\newcommand{\Tabref}[1]{Table~\ref{#1}}

\newcommand{\Figref}[1]{Fig.~\ref{#1}}

\DeclareMathAlphabet{\mathcal}{OMS}{cmsy}{m}{n}

\permission{Copyright is held by the International World Wide Web Conference Committee (IW3C2). IW3C2 reserves the right to provide a hyperlink to the author's site if the Material is used in electronic media.}
\conferenceinfo{WWW 2015,}{May 18--22, 2015, Florence, Italy.}
\copyrightetc{ACM \the\acmcopyr}
\crdata{978-1-4503-3469-3/15/05. \\
http://dx.doi.org/10.1145/2736277.2741666}

\begin{document}%

\title{
Mining Missing Hyperlinks from Human Navigation Traces:\\
A Case Study of Wikipedia
}

\numberofauthors{3}
\author{
\alignauthor
Robert West \\
\affaddr{Stanford University}\\
\email{west@cs.stanford.edu}
\alignauthor
Ashwin Paranjape \\
\affaddr{Stanford University}\\
\email{ashwinp@cs.stanford.edu}
\alignauthor
Jure Leskovec \\
\affaddr{Stanford University}\\
\email{jure@cs.stanford.edu}
}

\maketitle
\begin{abstract}

Hyperlinks are an essential feature of the World Wide Web. They are especially important for online encyclopedias such as Wikipedia: an article can often only be understood in the context of related articles, and hyperlinks make it easy to explore this context. But important links are often missing, and several methods have been proposed to alleviate this problem by learning a linking model based on the structure of the existing links. 

Here we propose a novel approach to identifying missing links in Wikipedia. We build on the fact that the ultimate purpose of Wikipedia links is to aid navigation. Rather than merely suggesting new links that are in tune with the structure of existing links, our method finds missing links that would immediately enhance Wikipedia's navigability. We leverage data sets of navigation paths collected through a Wikipedia\hyp based human\hyp computation game in which users must find a short path from a start to a target article by only clicking links encountered along the way. We harness human navigational traces to identify a set of candidates for missing links and then rank these candidates. Experiments show that our procedure identifies missing links of high quality.
\end{abstract}






\section{Introduction}
\label{sec:intro}

\noindent
The success of the World Wide Web hinges on the hyperlinks that weave its many billions of documents together. It is this fact that gave rise to its very name.
Hyperlinks are essential for several reasons.
From a human--computer interface perspective, they allow users to explore the available information in a natural way by effortlessly following pointers to references. For instance, the act of clicking a Wikipedia link is negligible compared to the cumbersome effort of flipping through countless pages of a paper encyclopedia, or even finding another volume on yet another shelf in a physical library.
Links are also important from an information management perspective.
They are among the most prominent features used by search engines, both for indexing and ranking.
When building a Web index, anchor texts serve as informative descriptors of the target page they point to, oftentimes more so than page titles and content \cite{manning-et-al2008}.
Moreover, in search\hyp result ranking, several standard methods, such as PageRank \cite{brin+page1998} and HITS \cite{kleinberg1999authoritative}, rely on the Web graph induced by the hyperlinks. And finally, links are important from a content provider perspective, since they make content discoverable to users and search engines. If a document has no incoming links, it cannot be accessed by a browsing user, nor can it be crawled and indexed by a search engine.

\begin{figure}
 \centering
	\includegraphics[width=\columnwidth]{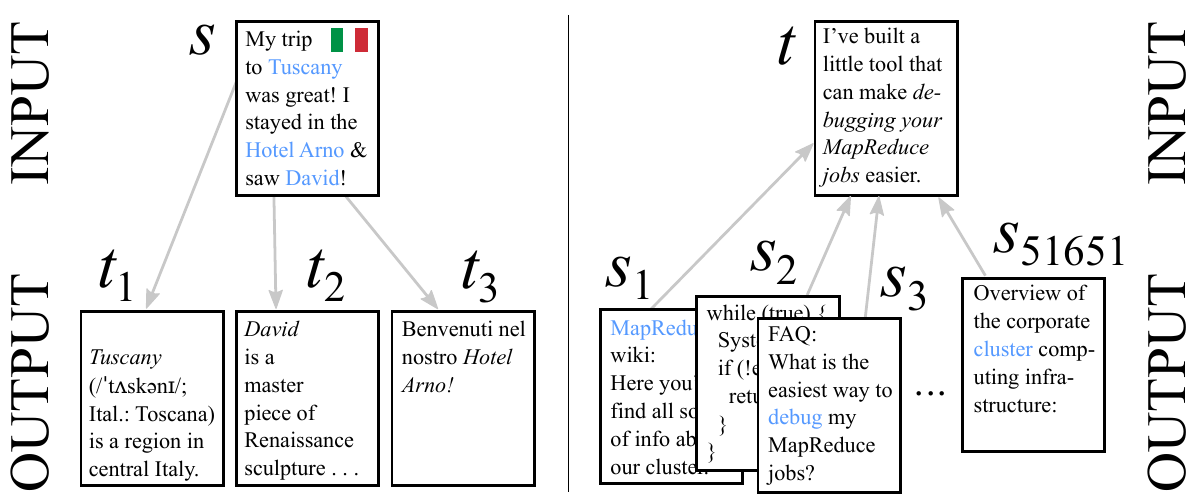}
	\hspace*{-2mm}
	(a) Target prediction
	\hspace{10mm}
	(b) Source prediction
	\caption{In scenario (a) a source $s$ is given, and the aim is to find mentions of relevant concepts in the source page and link them to appropriate targets. Here the set of candidate anchors is limited to the source document. In scenario (b) a target article $t$ is given, and the aim is to identify sources that contain relevant mentions of $t$ and could benefit from linking to it. Here every document mentioning the target is a potential source.}
 \label{fig:motivation}
\end{figure}

\xhdr{Source vs.\ target prediction}
While maintaining a good hyperlink structure is important, it is also difficult and time\hyp consuming.
We distinguish two tasks that typically arise during webpage maintenance, which we term \emph{target prediction} and \emph{source prediction:}

\begin{itemize}[noitemsep]
\item In the \emph{target prediction} task (\Figref{fig:motivation}(a)), a source document $s$ is given, and the goal is to find mentions of relevant concepts in $s$ and link them to appropriate targets $t$.
\item In the \emph{source prediction} task (\Figref{fig:motivation}(b)), a target document $t$ is given, and the goal is to identify sources $s$ that contain relevant mentions of $t$ and would benefit from referencing $t$.
\end{itemize}

To illustrate these two abstract tasks, consider the following concrete scenarios (schematized in \Figref{fig:motivation}).

\emph{Target prediction:}
In a typical target prediction scenario, a traveler might write a blog post about his recent trip to Tuscany, in which he mentions several places, foods, landmarks, and historical persons. To provide more context, he wants to link these mentions to external resources, such as Wikipedia articles, online recipes, or hotel websites. Manually identifying the relevant concepts and linking them to the most appropriate target pages can be a tedious process~\cite{devanbu1999chime,Nentwich:2002:XCC}.

{\emph{Source prediction:}
A typical source prediction scenario might involve a software engineer in a large company who has just finished a piece of code that could increase the productivity of many colleagues. She also created a documentation and tutorial page, but for it to be visible, she needs to link to it from other pages that colleagues interested in her code are likely to visit, such as company\hyp internal wikis, Q\&{}A fora, etc.
Along the same lines, consider a Wikipedia editor who has just written a new article. The article is of little use if it is not reachable from other articles, so the editor wants to plant links into other articles to point to the new article.

Identifying appropriate sources for a given target is even more difficult than identifying appropriate targets for a given source: In target prediction, the set of candidate anchors is limited and can be identified by inspecting the source document. In source prediction, on the contrary, the set of candidate sources is practically unbounded, as any page on the Web is a potential candidate to link to the target.}

\xhdr{Existing approaches} Automatic methods for detecting missing links would be important and useful.
Previous work has proposed methods mainly for the target prediction problem; \ie, they annotate a given source document with links to external resources, primarily Wikipedia articles.
One class of techniques can process arbitrary plain-text documents.
Here the text of the input document is combined with background knowledge from Wikipedia's textual content as well as graph structure to predict outgoing links \cite{meij2012adding,mihalcea+csomai2007,milne+witten2008_link}. A second group of approaches takes Wikipedia articles as input and uses the already existing links (and possibly the text content) to predict further outgoing links for the input article, \eg, based on adjacency\hyp matrix factorization \cite{west-et-al2009a,west-et-al2010}, information retrieval techniques \cite{adafre+derijke2005,wu+weld2007}, or machine learning \cite{noraset2014adding}.

With regard to the source prediction problem, a very simple approach would be to first collect all anchor texts in the document collection that frequently link to the target $t$ and to then link all as yet unlinked occurrences of these anchor texts to $t$ as well.
Unfortunately, this method is too simplistic and suffers from some major drawbacks: first, a phrase might not be link\hyp worthy in every context (\eg, `flower' is a good link anchor in the context of botany, but not in that of a wedding, where the concept of a flower needs no further explanation);
second, the disambiguation problem is not addressed
(\eg, `Florence' should link to \cpt{Firenze} in most contexts, but to \cpt{Florence, Alabama} in the context of Lauderdale County, Alabama).
Hence more sophisticated algorithms are required for solving the source prediction task.

Independent of whether the source or target prediction task is considered, what is largely missing from the picture is
the realization that, beyond text content and graph structure, there are additional sources of data that could be utilized in order to detect missing links more accurately. In particular, one such source of data that has remained mostly unexplored is human navigational traces on websites. Such traces are captured by the usage logs recorded on the server side by many websites, and the question arises: How can human click trails be harnessed in order to detect missing links?

\xhdr{Present work: Navigation logs for mining missing links}
In this paper we explore the use of human navigational traces for detecting missing links in websites like Wikipedia.
Logs of website navigation contain strong signals with regard to which existent hyperlinks are useful: from a user interface perspective, if a link is traversed often by humans then it is useful, and if it is never traversed by humans then it is redundant (although from a search engine's perspective it could still be useful for indexing and ranking). Hence we expect such logs to also contain clues into whether an as yet non\hyp existent link \emph{should} be there or not. For instance, if we often observe users going through page $s$ and ending up in page $t$, although $s$ does not directly link to $t$, then it might be a good idea to introduce a `shortcut' link from $s$ to $t$.
  
\begin{figure}
 \centering
	\includegraphics[width=\columnwidth]{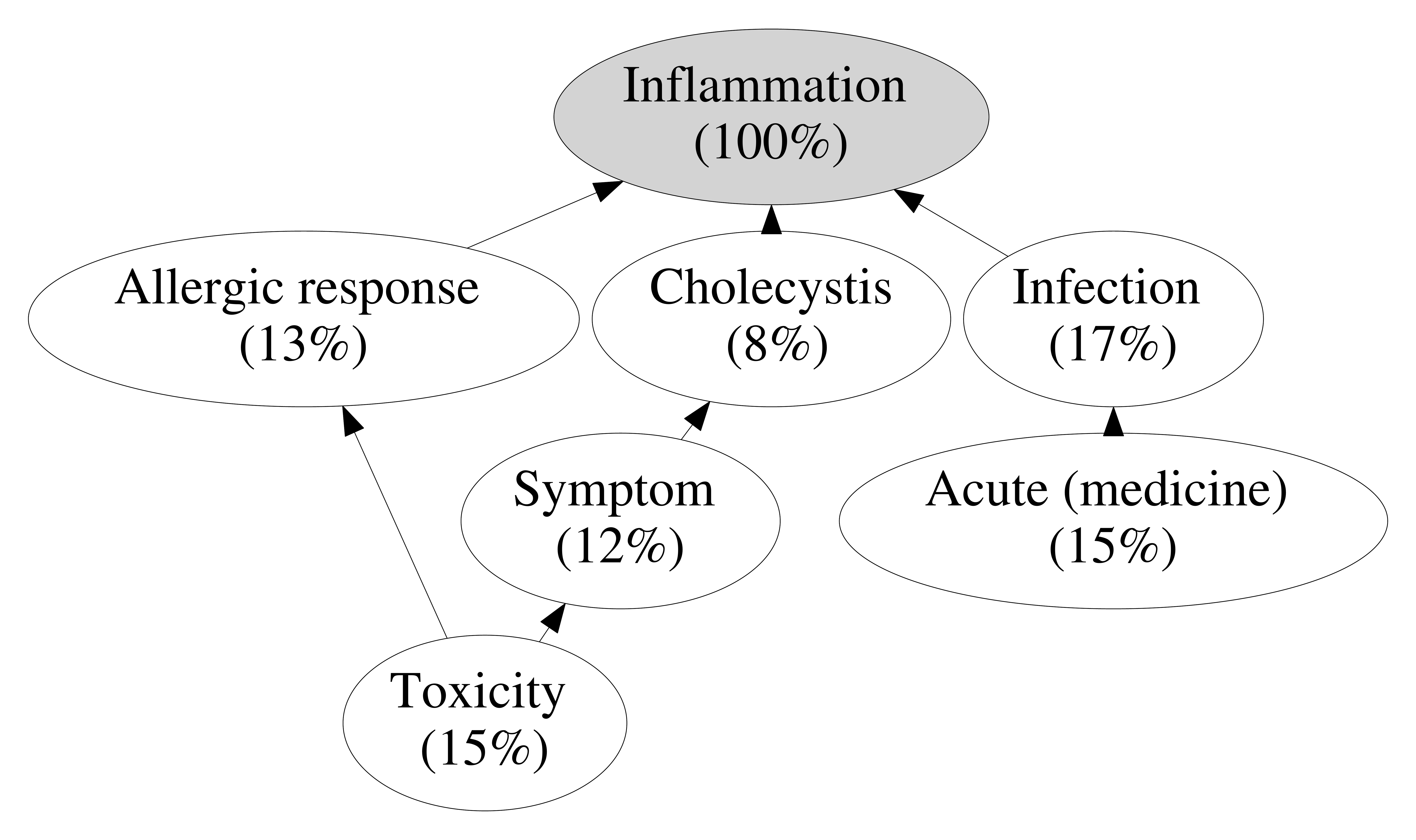}
	\caption{
The final portions of several navigation paths with the same target $t=$ \cpt{Inflammation}. The unfilled nodes are Wikipedia articles that appeared on paths to $t$. The number in each node indicates the percentage of paths with target $t$ that passed through that node.
}
 \label{fig:logarithm_example}
\end{figure}

As an analogy, consider the task of improving a road network.
A civil engineer would not just look at the existing road segments and try to infer which road segments to build next.
Rather, she would take into account how heavily each road segment is used and would then decide where it would make sense to add a shortcut, an extra lane, or a traffic light.
We argue that similarly we should consider how the Web's hyperlink structure is used and decide on that basis what hyperlinks to add next.
The \textit{raison d'\^{e}tre} of hyperlinks is to enable navigation, so by creating hyperlinks that aid navigation, we are optimizing the right objective.

\xhdr{Proposed approach to source prediction}
Here we propose a method for using navigational data to discover missing links following the above intuitions, thereby addressing the source prediction problem; \ie, given a target page, we find good sources to link to the target.
We demonstrate the effectiveness of our approach using Wikipedia.
We chose Wikipedia as our proof\hyp of\hyp concept domain because high\hyp quality navigation logs are available for it, collected via a class of human\hyp computation games known under names such as \textit{The Wiki Game} \cite{thewikigame} or \textit{Wikispeedia} \cite{wikispeedia,west-et-al2009}.
In these games, users are given two Wikipedia articles---a \textit{start} and a \textit{target}---and need to find a short path from the start to the target by traversing links encountered in the visited articles.
The underlying graph structure is unknown to users; they only see the outgoing links of the page they are currently on. But, crucially, they also have expectations on which pages should link to which other pages, based on their commonsense and expert knowledge about the world, and are guided by these expectations toward articles they consider likely to contain links to the target.

We consider our approach to be general and applicable to websites other than Wikipedia. Also note that we 
simply use navigational traces from the Wikipedia games since they are readily available to us. Obtaining 
raw, passively collected browsing logs of Wikipedia is much harder due to privacy considerations. However, we are encouraged to believe that our approach will generalize to passively collected browsing logs as well.

Building on the above intuition about humans browsing the Web, we reason as follows: if page $s$ is traversed by many users in search of target $t$, then this is an indicator that users expect the link from $s$ to $t$ to exist. So if $s$ does not link to $t$ yet (or not any more, for that matter), but contains a phrase that could be used as an anchor for $t$, then we should consider the link $(s,t)$ for addition.

As a concrete example, consider \Figref{fig:logarithm_example}.
The figure summarizes several navigational paths, all with the target $t=$ \cpt{Inflammation}.
Paths progress from bottom to top, and only the last few clicks are shown per path.
Each node $s$ also contains the fraction of all paths with target \cpt{Inflammation} that passed through $s$. 
For instance, we see that 17\% of times \cpt{Inflammation} was reached from \cpt{Infection} and 13\% of times it was reached from \cpt{Allergic response}.
A considerable fraction of paths (15\%) passed through \cpt{Acute (medicine)}, which does not link to $t$, although it mentions $t$ several times and could clearly benefit from a link to it. 

The central part of our approach is that we mine
many link candidates $(s,t)$
from a large number of navigation traces for each target $t$
and then rank these candidates by relevance.

We perform a set of experiments using automatically (and thus only approximately) defined ground-truth missing links, as well as an evaluation involving human raters.
In our automatically defined ground truth, we consider as positive examples of missing links such links that existed for a substantial amount of time but are missing from the latest Wikipedia snapshot.
In our evaluation by humans, raters labeled the identified missing links as relevant or not. Experiments show that restricting the candidate set to pairs observed in paths and then ranking those candidates using a simple heuristic performs better than applying more sophisticated ranking methods to the set of all possible candidates (\ie, including those not observed in paths). The reason why simple ranking methods suffice is that the `heavy lifting' is done by the users before ranking, by using vast amounts of world knowledge to select the pages that are best\hyp suited to link to the target.

The remainder of the paper is structured as follows.
\Secref{sec:method} provides details about the navigation trace data set and our methods for candidate selection and ranking.
We present experimental results in \Secref{sec:evaluation} and provide further discussion and perspectives in \Secref{sec:discussion}.
Related work is reviewed in \Secref{sec:relatedwork}, and \Secref{sec:conclusions} concludes the paper.

\section{Using human navigation logs\\for mining missing links}
\label{sec:method}

\begin{figure*}
 \centering
 \includegraphics[width=\textwidth]{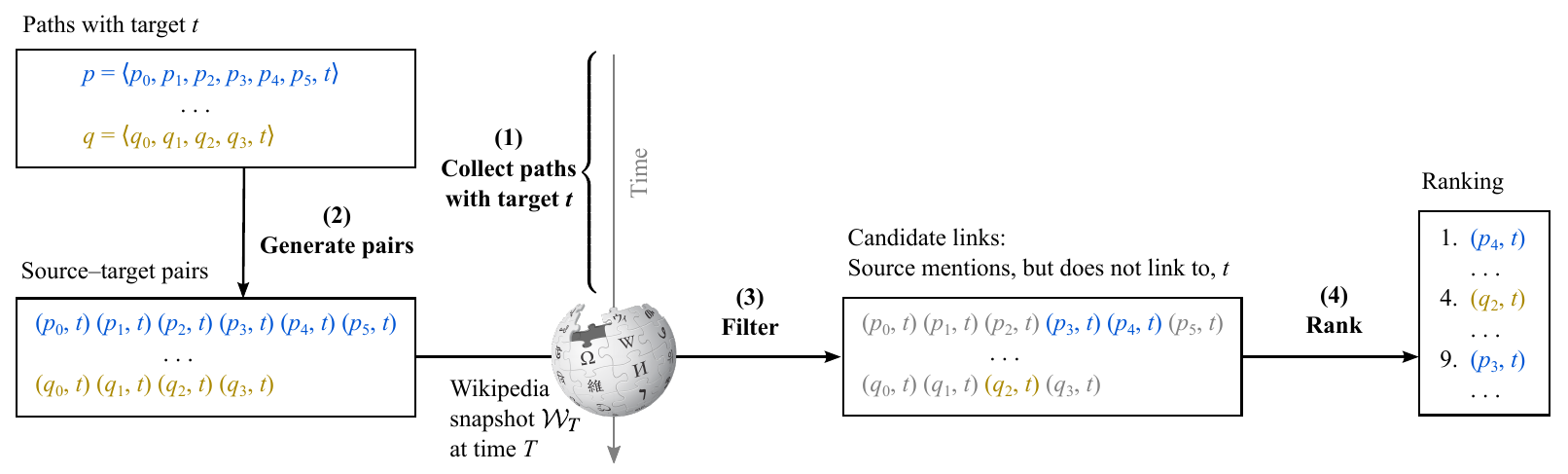}
 \caption{Overview of our approach for mining missing hyperlinks to a given target $t$ from human navigation traces. (1)~Collect paths with target $t$ up to time $T$, and capture the reference Wikipedia snapshot $\mathcal{W}_T$ at time $T$. (2)~Generate source--target pairs. (3)~Filter the pairs based on $\mathcal{W}_T$: a pair $(s,t)$ becomes a candidate if $s$ mentions, but does not link to, $t$ in $\mathcal{W}_T$; also exclude $(s,t)$ if $s$ tends to appear in the second half of paths with target $t$. (4)~Rank the candidate links.}
	 \label{fig:pipeline}
\end{figure*}

\noindent
In this paper we address the source prediction task of \Figref{fig:motivation}(b):
Given a target page $t$, we suggest a good list of source pages $s$ that should link to $t$.
Although this paper is focused on the special case of Wikipedia, we believe that our approach is general enough to be applied to other websites beyond Wikipedia. We discuss implications and requirements of this extension in \Secref{sec:discussion}.

We selected Wikipedia as our domain because, first, it constitutes an important and relevant special case due to its reliance on the links between articles, and because, second, high\hyp quality navigation traces are available for it.
In this section we first describe how these data were collected and then give a detailed account of how we use them for source prediction.

\subsection{Data sources}
\label{sec:the_wiki_game}

\noindent
Our data sets of Wikipedia navigation traces were collected via a popular online game that is generically known as `Wikiracing'~\cite{wikipedia:wikiracing}.
Several websites offer versions of this game, such as \textit{The Wiki Game} \cite{thewikigame} or \textit{Wikispeedia} \cite{wikispeedia,west-et-al2009}, but they all share the same general idea:
a user is given two Wikipedia articles---a \textit{start} and a \textit{target}---and is asked to navigate from the start to the target by exclusively clicking hyperlinks contained in the visited pages.
We also refer to start--target pairs as \textit{missions}.
In our experiments we use data from both The Wiki Game (\secref{sec:Evaluation The Wiki Game}) and Wikispeedia (\secref{sec:Evaluation Wikispeedia}).
Before describing how exactly we do so, we provide more details about the two data sets.

\xhdr{The Wiki Game}
In The Wiki Game, users may choose from five challenges:
`least clicks' (minimize the number of clicks),
`speed race' (minimize time),
`five clicks [or fewer] to Jesus' (find \cpt{Jesus} in five or fewer clicks, minimizing time),
`no United States' (minimize time while avoiding the \cpt{USA} article),
and `six degrees of Wikipedia' (minimize time while finding a path of exactly six clicks).
In each challenge, several players compete for the same mission simultaneously, navigating the full Wikipedia.
We pool the paths collected from all five challenges between 2009 and 2012, thus obtaining a data set of 974k paths grouped into 364k distinct missions (start--target pairs); \ie, there are 2.7 paths per mission on average.
The number of distinct targets is 3,219, \ie, we have 303 paths per target on average, with a median of 208.
Targets with many paths are quite frequent; \eg, there are 2,087 targets with at least 100, and 698 targets with at least 500, paths.
In the following, we focus on the 2,087 targets (65\% of all targets) that have at least 100 paths.


\xhdr{Wikispeedia}
Unlike The Wiki Game, Wikispeedia is a single\hyp player game.
Once a mission is successfully completed, the user may enter her name into a high\hyp score table associated with that mission, where users are ranked by number of clicks, with ties broken by time.
The game is played on a reduced, static snapshot of Wikipedia containing 4,604 of the most important articles \cite{schools-wikipedia}.
The data set we work with is publicly available \cite{wikispeedia-data} and comprises 51k paths collected from 2009 to 2014, grouped into 29k distinct missions, for an average of 1.8 paths per mission.
The number of distinct targets is 3,326; \ie, we have 15 paths per target on average, with a median of 10.

Comparing the two data sets, we notice that The Wiki Game, on the one hand, has the advantage of more data, in particular more paths per target.
Wikispeedia, on the other hand, has the advantage of using a static Wikipedia snapshot (while The Wiki Game fetches pages from live Wikipedia on the fly and caches them for some time \cite{clemesha2014}), which allows for a different kind of evaluation (\Secref{sec:Evaluation Wikispeedia}), and of being publicly available, which makes our experiments reproducible.

\subsection{Source candidate selection}
\label{sec:candidate_selection}

\noindent
Our method for source prediction consists of four steps, as illustrated in \Figref{fig:pipeline}. (The fourth step, ranking, is covered in \Secref{sec:candidate_ranking}.)

\xhdr{Step 1: Collect paths}
We start by collecting navigation paths with the given target $t$, up to time $T$.

\xhdr{Step 2: Generate pairs}
For each path $p = \langle p_0, \dots, p_n=t \rangle$, the initial set of candidates is $\{ (p_i,t) \; : \; 0 < i < n \}$, \ie, every direct link from any page along the path to the target $t$ is initially a candidate. (The start page $p_0$ is exempt, since it is selected randomly and is therefore unlikely to be a useful candidate.)
There are in general many paths for the same target $t$ (upper left box in \Figref{fig:pipeline}), so we take the union of the candidate sets resulting from all these paths (lower left box) as the initial candidate set for $t$.

\xhdr{Step 3: Filter}
Next, we filter this initial set using the Wikipedia version $\mathcal{W}_T$ at time $T$, which serves as our \emph{reference snapshot}.
A link $(s,t)$, where $s \in \{p_1, \dots, p_{n-1}\}$, can be suggested only if it does not already exist in $\mathcal{W}_T$.
Further, the source $s$ should contain a phrase that could serve as the anchor for a link to $t$; in other words, $s$ should mention $t$ in $\mathcal{W}_T$.

To detect pages that mention the target $t$, we construct the set $\mathcal{A}_t$ of all phrases that serve as anchor texts for $t$ across all articles in the reference Wikipedia snapshot%
\footnote{In practice, we exclude (1)~phrases that rarely (less than 6.5\% of all cases \cite{milne+witten2008_link}) serve as link anchors for any target, which excludes, \eg, `A' as an anchor for \cpt{Ampere}, and (2)~anchor texts for which $t$ is seldom (less than 1\% of all cases) the target, which excludes, \eg, `Florence' as an anchor for \cpt{Florence, Alabama}.} and subsequently define that $s$ mentions $t$ if it contains any phrase from $\mathcal{A}_t$.

Previous work \cite{west+leskovec2012www} has shown that navigation traces tend to consist of a `getting\hyp away\hyp from\hyp the\hyp start' phase, in which the user attempts to reach a hub article that is well\hyp connected in the network of pages, and a `homing\hyp in\hyp on\hyp the\hyp target' phase, in which the user actively seeks out pages related to $t$.
Guided by this finding, we apply one additional filtering step and include in our final candidate set (the box labeled `candidate links' in \Figref{fig:pipeline}) only sources that tend to appear in the second half of paths with target $t$.
More precisely, we first define the \emph{relative path position} of $p_i$ along the path $p = \langle p_0, \dots, p_n=t \rangle$ to be $i/n$, and then discard the pair $(s,t)$ if the relative path position of $s$ on paths with target $t$ is less than or equal to 0.5 on average.

\subsection{Source candidate ranking}
\label{sec:candidate_ranking}

\noindent
Source candidate selection yields an unordered set of candidates for each target $t$.
The goal of the next (and final) step in our pipeline is to turn this set into a meaningful ranking (step~4 in \Figref{fig:pipeline}).
Since the source prediction task (\Figref{fig:motivation}(b)) asks for sources for a given target $t$, we produce a separate ranking for each $t$.
Several ranking methods are conceivable:

\begin{enumerate}
\item \textbf{Ranking by relatedness.} It seems reasonable to rank source candidates $s$ by their relatedness to $t$, since clearly a link is more relevant between articles with topical connections.%
\footnote{According to the Wikipedia linking guidelines \cite{wikipedia:linking}, links should correspond to `relevant connections to the subject of another article that will help readers understand the article more fully.'}
Since we deal with Wikipedia as our data set, we choose relatedness measures based on Wikipedia (see below).
\item \textbf{Ranking by path frequency.} Navigation traces provide us with statistics about how frequently a source $s$ was traversed by users searching for target $t$. Based on this, we compute the \textit{path frequency} of $s$ given target $t$, defined as the fraction of paths that passed through $s$, out of all the paths with target $t$.
Intuitively, pages $s$ that were traversed more frequently on paths to $t$ should be better sources for links to $t$.
\end{enumerate}

We experiment with two relatedness measures for case~1 above.
The first is due to Milne and Witten \cite{milne+witten2008} and is based on the inlink sets $\mathcal{S}$ and $\mathcal{T}$ of $s$ and $t$, respectively.
It calculates the distance between $s$ and $t$ as the negative log probability of seeing a link from $\mathcal{S} \cap \mathcal{T}$ when randomly sampling a link from the larger one of the sets $\mathcal{S}$ and $\mathcal{T}$ (normalized to approximately lie between~0 and~1), and the relatedness as one minus that distance:
\begin{equation}
\operatorname{MW}(s,t) = 1 - \frac{\log(\max\{|\mathcal{S}|, |\mathcal{T}|\}) - \log(|\mathcal{S} \cap \mathcal{T}|)}{\log(N) - \log(\min\{|\mathcal{S}|, |\mathcal{T}|\})},
\label{eqn:MW}
\end{equation}
where $N$ is the total number of Wikipedia articles.
	
The second relatedness measure is due to West \etal\ \cite{west-et-al2009a} and works by finding a low-rank approximation of Wikipedia's adjacency matrix via the singular\hyp value decomposition (SVD). The pair $(s,t)$ corresponds to an entry $A[s,t]$ in the adjacency matrix $A$ and to an entry $A_k[s,t]$ in the rank-$k$ approximation $A_k$ obtained from $A$ via SVD. If $A[s,t]=0$ and $A_k[s,t] \gg 0$ then $s$ does not link to $t$ yet but is a good candidate. Therefore we define the SVD-based relatedness as
\begin{equation}
\operatorname{SVD}(s,t) = A_k[s,t] - A[s,t].
\label{eqn:SVD}
\end{equation}
In our experiments on The Wiki Game, we use the reduced rank $k=$ 1,000.
Since the adjacency matrix is much smaller for Wikispeedia (\Secref{sec:the_wiki_game}), we use the smaller value of $k=256$ there.

\subsection{Exploratory analysis of link candidates}
\label{sec:exploratory_analysis}

\noindent
Having introduced the data set and our source prediction method, we now explore the data set of human navigation traces to build intuitions on strengths and potential weaknesses of our approach.

We use the Wikipedia version as of $T=$ 2014-01-02 as our reference snapshot $\mathcal{W}_T$ in all experiments.

\xhdr{Number of pages on a path mentioning the target}
We count for each path $p = \langle p_0, \dots, p_n=t \rangle$ how often the target $t$ is mentioned across all visited nodes $p_1, \dots, p_{n-1}$ (excluding the randomly selected start page $p_0$) and find that, on average, $t$ is mentioned on 1.7 pages per path.
Since $p_{n-1}$ contains a link to $t$, it is very likely to also mention $t$ (for our definition of a mention, \cf\ \Secref{sec:candidate_selection}), which means that, on average, each path contains 0.7 additional pages that mention $t$.

Now consider the subset of visited pages that mention $t$. Out of these, 73\% contain a link to $t$ in the reference Wikipedia snapshot $\mathcal{W}_T$.
The remaining 27\%, which do not link to $t$ in $\mathcal{W}_T$, are potentially good candidate sources to link to $t$, since these pages were actively chosen by the user while  searching for $t$.



\xhdr{Properties of pages along paths}
We also investigate which parts of a path carry most value for source prediction.
Consider \Figref{fig:prob_mention}, which aggregates all paths and shows for each part of the path how likely the pages in that part are to mention $t$.
In order to be able to aggregate paths of variable length, we adopt the notion of relative path position (\Secref{sec:candidate_selection}).
\Figref{fig:prob_mention} uniformly buckets the range $[0,1]$ into five intervals and plots the average for each interval.
We only include paths of at least five clicks, such that each path contributes to each bucket, and the page $p_{n-1}$ just before the target always falls into the last bucket.

We see that target mentions become more frequent as paths progress (the black curve in \Figref{fig:prob_mention}): two-thirds of pages with relative path positions in the interval $[0.6, 0.8)$ mention the target, while at positions in $[0.8, 1.0)$ nearly all pages (91\%) do.
We are particularly interested in mentions that are not accompanied by a link to $t$ (the magenta curve in \Figref{fig:prob_mention}), since these are our source candidates.
The figure tells us that candidates are more likely to appear towards the end of paths:
at relative path positions in the interval $[0.6, 0.8)$, 30\% of pages without a link to $t$ mention $t$, and for the interval $[0.8, 1.0)$, the fraction is as high as 46\%.

We note that these curves are in tune with previous work \cite{west+leskovec2012www}, which has shown that humans tend to follow a `semantic gradient' during information network navigation, passing through articles that get ever more related to the target. In this light it makes a lot of sense that the rate of target mentions should increase as paths progress.

\subsection{Obtaining ground truth based on Wikipedia evolution}
\label{sec:weak_ground_truth}

\noindent
In order to form intuitions about how meaningful our suggestions are, we would ideally like to evaluate for each relative path position how good the source candidates at that position are.
However, ground-truth data is hard to come by; in order to make strong claims, we need to ask humans how good our predictions are.
We do so later on (\Secref{sec:Evaluation based on human raters}), but since obtaining human ratings is expensive and time\hyp consuming, we preliminarily adopt a notion of ground truth that is approximate and biased, but nevertheless allows us to gain some initial insights.
In this subsection, we define this approximate ground truth and analyze our navigation traces in terms of it.

\begin{figure}[t]
	\centering
	\hspace{-3mm}
 	\subfigure[Fraction of target mentions]{
		\includegraphics{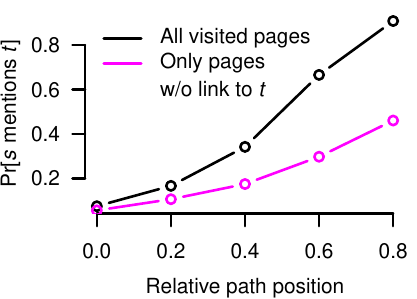}
		\label{fig:prob_mention}
	}
	\hspace{-3mm}
 	\subfigure[Fraction of good candidates]{
		\includegraphics{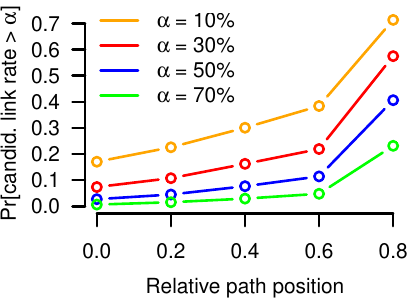}
		\label{fig:prob_good_link_given_mention}
	}
\caption{
\textbf{(a)}~Fraction of sources $s$ mentioning target $t$, as a function of the relative path position of $s$; the magenta curve is conditioned on $s$ not linking to $t$.
\textbf{(b)}~Fraction of candidates that are positive according to the automatically generated ground truth, as a function of relative path position, for several link-rate thresholds $\alpha$ (\Secref{sec:weak_ground_truth}); source candidates mention, but do not link to, the target and are considered positive if their link rate is greater than $\alpha$.
}
\vspace{-3mm}
\end{figure}

We obtain a weak notion of ground truth from the evolution of Wikipedia's graph structure as follows.
First we define the \emph{link rate} of $(s,t)$ as the fraction of time $s$ contained a link to $t$ since $s$ was created.
(We compute these values based on Wikipedia's complete edit history.)
Then we choose a \emph{link-rate threshold} $\alpha \in [0\%, \dots, 100\%]$ and label a candidate link $(s,t)$ as positive if its link rate is greater than $\alpha$.
Candidates with a positive label correspond to links that existed for a substantial amount of time, but got deleted before the reference Wikipedia snapshot $\mathcal{W}_T$ (\cf\ step~3 of \Secref{sec:candidate_selection}).
That is, such links could have been valuable for navigation, yet were removed at some point in time, so we argue that reintroducing them is likely to improve Wikipedia.

Consider a candidate $(s,t)$ labeled as positive according to the above definition.
The link $(s,t)$ may (case~1), or may not (case~2), have existed during the game from which it was mined. Further, if it existed during the game, it may (case~1a), or may not (case~1b), have been clicked by the user.
These three cases correspond to the following scenarios.
If the link $(s,t)$ existed during the game and was clicked by the user (case~1a), but has been deleted since (as required for $(s,t)$ to be a candidate), then it is probably a good idea to suggest it for reintroduction.
If the link existed during the game, but was not clicked by the user (case~1b), this means that she did not see it in her rush to reach $t$ as fast as possible (or else she would have clicked on it to immediately win the game); so either the user found another promising way to continue the search before seeing the link to $t$, or the link was too hard to find in the text of $s$, which is a signal that we should reintroduce that link and make it more obvious.
Finally, if $(s,t)$ did not exist during the game (case~2) then the user could not possibly have taken it, although she might have intended to do so (since she actively navigated to $s$ while searching for $t$); in this case, too, $(s,t)$ might be a good link suggestion.

We refer to our automatically obtained labels as `weak' because, by definition, they contain many \emph{false negatives}.
Wikipedia is an evolving organism, and an important part of our task is to suggest links which never existed. However, by the above link-rate threshold criterion, these links will be counted as negative examples.
For an example of such a false negative, consider again \Figref{fig:logarithm_example}, where the article on \cpt{Acute (medicine)} should clearly link to \cpt{Inflammation}, as it explains a concept critical to understanding the term \cpt{Acute} as used in medicine, but the link from \cpt{Acute (medicine)} to \cpt{Inflammation} is labeled as negative by the automatic ground truth, since \cpt{Acute (medicine)} has never linked to \cpt{Inflammation} in Wikipedia's history.
In other words, the automatically obtained ground truth has high precision, but low recall of truly positive examples.
Nonetheless, this weak ground truth is useful during development because it provides us with many labeled examples for free and allows for relative comparisons between different methods.

\Figref{fig:prob_good_link_given_mention} captures this approximate notion of candidate quality, again broken up by relative path position.
The graph shows that the fraction of positives obtained from the automatically obtained ground truth becomes higher for pages appearing later on in paths.
We conclude that not only are mentions at later positions more frequent (\Figref{fig:prob_mention}), but that they also correspond to better link anchors.
(We try several values for the link-rate threshold $\alpha$, but the same trend holds for all thresholds.)
This provides additional justification for our decision to include in our set of source candidates only sources that tend to appear in the second half of navigation traces with the given target (\Secref{sec:candidate_selection}).

\section{Evaluation}
\label{sec:evaluation}

\noindent
In our experiments we compare five methods:
Given a target~$t$, we can either consider as source candidates the set of \emph{all} articles that mention $t$ but do not link to it (across our entire reference Wikipedia snapshot $\mathcal{W}_T$);
or we can subselect candidate sources based on whether we observe them in navigation paths (\Secref{sec:candidate_selection}).
Further, we consider two relatedness measures for ranking (\Secref{sec:candidate_ranking}).
This yields four combinations of candidate selection methods (`none' and `path-based') and relatedness measures (`MW' and `SVD').
The fifth method requires no external relatedness measure but simply ranks candidates with respect to their frequency among paths with target $t$ (\Secref{sec:candidate_ranking}).

To sum up, we consider the following five methods for predicting missing links to a given target page $t$:
\begin{itemize}
	\item {\bf No selection, rank by MW:} Use all candidate sources and rank them based on the MW method (\Eqnref{eqn:MW}).
	\item {\bf No selection, rank by SVD:} Use all candidate sources and rank them based on the SVD method (\Eqnref{eqn:SVD}).
	\item {\bf Path-based selection, rank by MW:} Only use candidates appearing in navigational traces and rank them based on the MW method.
	\item {\bf Path-based selection, rank by SVD:} Only use candidates appearing in navigational traces and rank them based on the SVD method.
	\item {\bf Path-based selection, rank by frequency:} Only use candidates appearing in navigational traces and rank them based on the frequency with which they appear in paths (\cf\ \Secref{sec:candidate_ranking} for our definition of \textit{path frequency}).
\end{itemize}

\begin{figure*}[t]
	\centering
	\hspace{-1mm}
 	\subfigure[The Wiki Game\hspace{-5mm}]{
	\includegraphics{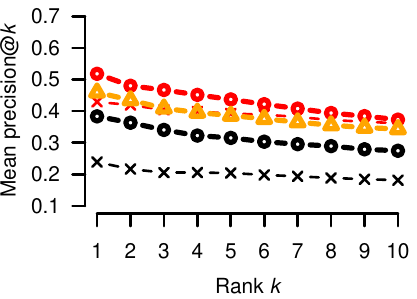}
		\label{fig:prec_at_k}
	}
	\hspace{-1mm}
 	\subfigure[The Wiki Game\hspace{-5mm}]{
	\includegraphics{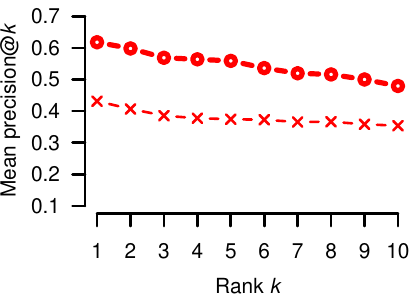}
		\label{fig:prec_at_k_mturk}
	}
	\hspace{-1mm}
 	\subfigure[Wikispeedia\hspace{-4mm}]{
	\includegraphics{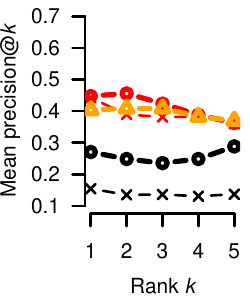}
		\label{fig:prec_at_k_wikispeedia_current}
	}
	\hspace{-1mm}
 	\subfigure[Wikispeedia\hspace{-4mm}]{
	\includegraphics{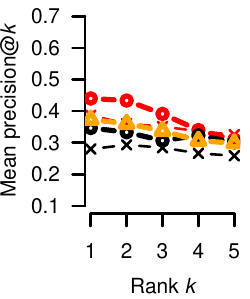}
		\label{fig:prec_at_k_wikispeedia_schools}
	}
	\includegraphics{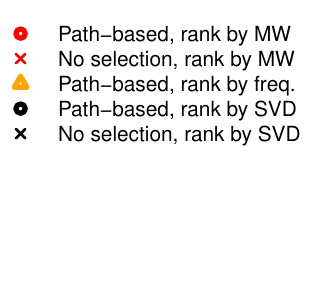}
\caption{
Performance in terms of precision@$k$ for different source selection and ranking methods on our two data sets. Bold lines represent path\hyp based candidate selection.
(a)~The Wiki Game, automatically obtained ground truth (\Secref{sec:Evaluation using automatically obtained ground truth}).
(b)~The Wiki Game, human ground truth (\Secref{sec:Evaluation based on human raters}); only the MW ranking method was used in the human evaluation. Note that performance as evaluated by humans exceeds the estimate from the automated evaluation (\Figref{fig:prec_at_k}), \ie, the latter underestimates the actual quality of suggested links.
(c)~Wikispeedia, automatically obtained ground truth, same evaluation methodology as applied to The Wiki Game (`standard evaluation'; \Secref{sec:Evaluation Wikispeedia}).
(d)~Wikispeedia, automatically obtained ground truth, stricter evaluation (\Secref{sec:Evaluation Wikispeedia}).
}
\label{fig:prec_at_k_ALL}
\end{figure*}

\xhdr{Ground truth}
We perform a twofold evaluation, one based on the automatically obtained and approximate labels defined in \Secref{sec:weak_ground_truth} (we use the link-rate threshold $\alpha=30\%$ throughout), the other based on labels obtained from human raters.
For our human evaluation, we select a subset of targets, predict sources for them using the methods that performed best during the development phase on the automatic ground truth, and ask raters on Amazon Mechanical Turk~\cite{mturk} to label the top predictions. Here we get rid of the shortcomings of the automatic ground truth, on which we cannot obtain absolute performance numbers (mainly due to the high false\hyp negative rate; \Secref{sec:weak_ground_truth}), but have less data to work with.

We perform an evaluation by humans only on the predictions obtained on data from The Wiki Game.

\xhdr{Evaluation metric}
As our evaluation metric, we use precision@$k$ for $k=1,\dots,K$.
We first calculate the $K$ precision values for each target separately and then compute the aggregate value for each $k$ by averaging over all targets.
This means we can only include targets for which our methods find at least $K$ source candidates (which naturally shrinks the set of test targets).
We use $K=10$ for The Wiki Game, which defines our evaluation set of 699 targets.
Since the Wikispeedia data set contains fewer paths, we are less restrictive here and choose $K=5$, obtaining an evaluation set of 181 targets.

The bulk of our experiments is performed in \Secref{sec:Evaluation The Wiki Game} on data from The Wiki Game.
Subsequently, \Secref{sec:Evaluation Wikispeedia} completes the evaluation by demonstrating that our algorithm works equally well on Wikispeedia.

\subsection{Evaluation on The Wiki Game}
\label{sec:Evaluation The Wiki Game}

\noindent
We start by evaluating our algorithm on data from The Wiki Game, first based on the automatically obtained ground truth, then by asking human raters.

\subsubsection{Evaluation using automatically obtained\\ground truth}
\label{sec:Evaluation using automatically obtained ground truth}

\noindent
The precision@$k$ curves for all five methods as evaluated on The Wiki Game are displayed in \Figref{fig:prec_at_k}, and their performance is summarized in terms of the area under the precision@$k$ curve in \Tabref{tbl:wikipedia_ground_truth}.

Overall, we achieve good performance, especially given that our ground truth is of high precision but low recall, with many false negatives. Even though the precision@$k$ lies in the range between 0.4 and 0.5 (for path-based candidate selection and MW ranking), manual error inspection revealed that most suggested links make sense and are truly missing, and that, in fact, the Wikipedia community has simply never included these links into the Wikipedia graph so that they could have made their way into our ground truth (\eg, $(\text{\cpt{Acute (medicine)}}, \text{\cpt{Inflammation}})$ in \Figref{fig:logarithm_example}).

Comparing the different methods, we observe that path-based candidate selection performs better than doing no subselection for both relatedness measures used in ranking. Path-based selection improves performance by a particularly large margin for the SVD-based ranking method, which has much lower precision@$k$ than the other methods. This establishes the fact that there is a lot of value in path-based candidate selection especially when the ranking measure does not excel by itself.

\begin{table}[tb]
	\centering
	\begin{tabular}{r|ccc}
	Candidate	& Rank by		& Rank by		& Rank by \\
	selection	& MW 	& SVD	& path freq. \\
	\hline
	None			&	39\% 	&	20\%		& N/A \\
	Path-based	&	43\%  	&	32\%		& 39\%    \\
	\end{tabular}
	\caption{Area under the precision@$k$ curve for no candidate selection versus path-based candidate selection for all ranking measures (The Wiki Game; \Figref{fig:prec_at_k}).
	Note that path frequency is only applicable for path-based candidate selection.}
	\label{tbl:wikipedia_ground_truth}
\end{table}

The margin between path-based selection and no selection is larger for smaller $k$, which means that considering navigational paths is particularly useful for predicting the top link sources.

Note that both relatedness measures (MW and SVD) use the high\hyp quality link structure of the Wikipedia page graph (\Secref{sec:candidate_ranking}).
If we wanted to generalize our approach to domains beyond Wikipedia, we can easily imagine scenarios where no such high\hyp quality relatedness measures are readily available (\eg, when pages are not as topically coherent as Wikipedia articles, or pages have scarce content and are poorly interlinked).
With such situations in mind, it is encouraging to see that our fifth measure (`rank by frequency' in combination with path\hyp based candidate selection; the yellow curve in \Figref{fig:prec_at_k}) performs quite competitively. Recall that that ranking method does not rely on any external relatedness measure but simply ranks source candidates with respect to the frequency with which they appeared on paths with target $t$.
This is an important observation because it means our method has the potential to generalize well to use cases where a good relatedness measure is not readily available.

\begin{figure}[t]
	\centering
	\hspace{-3mm}
 	\subfigure[Fraction of final clicks\hspace{-8mm}]{
	\includegraphics{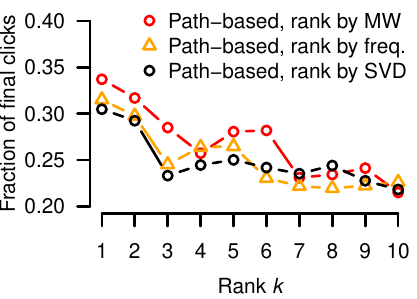}
		\label{fig:mean_isLastClick_at_k}
	}
	\hspace{-3mm}
 	\subfigure[Volume vs.\ precision\hspace{-6mm}]{
	\includegraphics{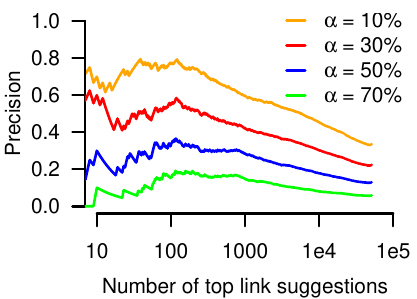}
		\label{fig:num_suggestions_vs_precision}
	}
\caption{
(a)~Fraction of clicks that were the final click along their respective path, as a function of rank.
If a suggestion corresponds to the final click along a path, the suggested link must have existed at game time, so a useful link is effectively reintroduced into Wikipedia by that suggestion.
(b)~Precision as a function of the total number of top link suggestions made by our method across all targets, where suggestions are ranked by frequency (\Secref{sec:candidate_ranking}). Positive suggestions are those whose link rate lies above the respective link-rate threshold $\alpha$ (\Secref{sec:weak_ground_truth}).
}
\end{figure}

\xhdr{Reintroduction of valuable but deleted links}
By construction, the last click on a path always leads into the target.
The fact that a user looked for, found, and clicked on this link is a very strong signal that the link is useful for navigation.
Removing such links from Wikipedia is particularly harmful from a user\hyp interface perspective, and it is desirable that a source prediction method suggest them for reintroduction.
To see if our path-based candidate selection method meets this desideratum, \Figref{fig:mean_isLastClick_at_k} plots, for each rank $k$, the fraction of predicted links that were also the last link on the paths they were mined from.
We observe that, while most suggested links were not clicked by humans (most likely because they were not present in the version of the Wikipedia page used by The Wiki Game), a substantial fraction (between 20\% and 35\%) correspond to links that existed at game time and were chosen by the user but do not exist in the reference snapshot any more.
We conclude that our top suggestions are often links that were taken by the user as the last click to the target but have since been removed, and thus our method rightfully reintroduces such links back into Wikipedia.

\xhdr{Total volume of added links}
So far we have conducted a per\hyp target evaluation, by first computing precision@$k$ values for each target and then averaging over all targets for each rank $k$.
But it is also interesting to consider the total number of links we can suggest at a given precision level, across all targets, since this gives us an idea of the potential number of improvements we could make to Wikipedia by deploying our system.
The results of this evaluation are presented in \Figref{fig:num_suggestions_vs_precision}, which shows that we can make 1,000 link suggestions at a precision of 42\%, and 10k suggestions at a precision of 30\% (ranking candidates by frequency and assuming the same link-rate threshold $\alpha=30\%$ used in our automated evaluation, corresponding to the red curve in \Figref{fig:num_suggestions_vs_precision}).

\subsubsection{Evaluation by human raters}
\label{sec:Evaluation based on human raters}

\noindent
Wikipedia is a continuously evolving entity.
Although the link history, on the basis of which we defined our automatic ground truth, captures this evolution, it can only tell us which links are positive examples (because they persisted throughout a long period of time).
However, there are many links that should be, but have never been, added to Wikipedia, and if our method suggests such a link, then the previous evaluation would count it as a bad suggestion.
Therefore the above notion of ground truth suffers from false negatives.
To combat this problem, we perform a more accurate evaluation by human raters in this section.
Having done so, we can also confirm the prevalence of false negatives \textit{post hoc} (see the end of this subsection).

\xhdr{Methods compared via human evaluation}
In our human evaluation, we compare two of the top\hyp performing methods:
(1)~path-based candidate selection with MW ranking
and (2)~no candidate selection with MW ranking.
By using the same ranking method and only switching whether path-based candidate selection was performed, we can gauge the impact of the latter on performance.


\xhdr{Target sampling}
In order to select targets on which to evaluate the predictions of the two methods, we stratify the base set of 699 targets by the number of paths observed per target and select ten targets from each decile, for a total of 100 test targets.
The rationale behind stratification is that we want to avoid being biased towards targets for which the path-based candidate selection method can make a particularly large number of good predictions (because more data are available for those targets).

\xhdr{Obtaining ratings through Amazon Mechanical Turk}
We use Amazon Mechanical Turk \cite{mturk} for recruiting human raters.
As in the automatic evaluation, our goal is to assess the precision@$k$, where $k=1,\dots,10$, for the two compared methods.
In each rating task, the human evaluator was presented with a target $t$ and a set of 14 candidate sources
and was asked to indicate
which of the candidates should contain a link to the target article.
There were no constraints on the number of source articles the rater could choose. 
The set of 14 candidate sources comprised the following entries:
\begin{enumerate}
	\item Five predictions from each of the two compared methods (either suggestions 1 through 5 or suggestions 6 through 10 from each method).
	\item Two control sources, sampled randomly from the set of all Wikipedia articles that link to the target $t$.
	\item Two control non-sources, sampled randomly from the set of all Wikipedia articles and hence highly unlikely to link to $t$.
\end{enumerate}

In cases where the two methods agreed on a suggestion, that suggestion was included only once in the set of source candidates, thereby making the presented list shorter than the maximum of 14 items.
Also, to prevent any ordering bias, we shuffled the order of sources in the presented list.
The task description is reproduced \textit{verbatim} in Appendix~\ref{app:instructions}.

We paid 5\textcent{} per task, and each task was presented to ten different workers.
We consider a source to be a positive example if over half of the ten raters labeled it as such.

\Figref{fig:prec_at_k_mturk} presents the results. We observe that path\hyp based candidate selection followed by MW ranking outperforms MW ranking on the set of all candidates by a large margin.
\Tabref{tbl:mturk_ground_truth}, which summarizes the performance of both compared methods on the human\hyp labeled ground truth (again as the area under the precision@$k$ curve) and compares it to the performance obtained on the automatically labeled ground truth, shows that the area under the precision@$k$ curves for path\hyp based candidate selection increases by 12\%, compared to the automatically obtained ground truth. 
On the other hand, when doing no candidate selection, the area under the curve decreases by 1\%.

The reasons for the increased performance on the human\hyp labeled ground truth are twofold.
First, the automatically obtained ground truth uses only a historical notion of correctness in which many actually positive examples are mislabeled as negative.
Second, MW relatedness alone, without performing path-based candidate selection, might not capture the notion of human\hyp intuition--based similarity well. 
Path-based selection, on the contrary, captures exactly that quality by design, and it is thus not surprising that it prevails on a human-labeled ground truth by such a large margin. 

Out of the controls that represent randomly selected sources already linking to the target page (item~2 in the above list of source\hyp candidate types presented to raters),
only 9\% are labeled as positive by more than five of the ten raters, a value much lower than even our precision@10 of about 50\%.
This tells us that the links we suggest are better than the average pre-existing link to the target.

Finally, out of the random control non-sources (item~3 in the above list), only one pair (\cpt{Geography of Korea} to \cpt{South Korea}) was rated positive by more than five of the ten raters (we happened to sample a connected pair here). This statistic confirms that human labeling was not random. 

\begin{table}[tb]
\centering
\begin{tabular}{r|cc}
Candidate	& Automatic		& Human-labeled\\
selection	& ground truth 			& ground truth		\\
\hline
None			&	39\% 	&	38\%\\
Path-based	&	43\%  	&	55\%\\
\end{tabular}
\caption{Area under the precision@$k$ curve for MW ranking, comparing the automated (\Figref{fig:prec_at_k}) and human (\Figref{fig:prec_at_k_mturk}) evaluations of our method run on data from The Wiki Game.}
\label{tbl:mturk_ground_truth}
\end{table}

\begin{figure}
 \centering
	\includegraphics{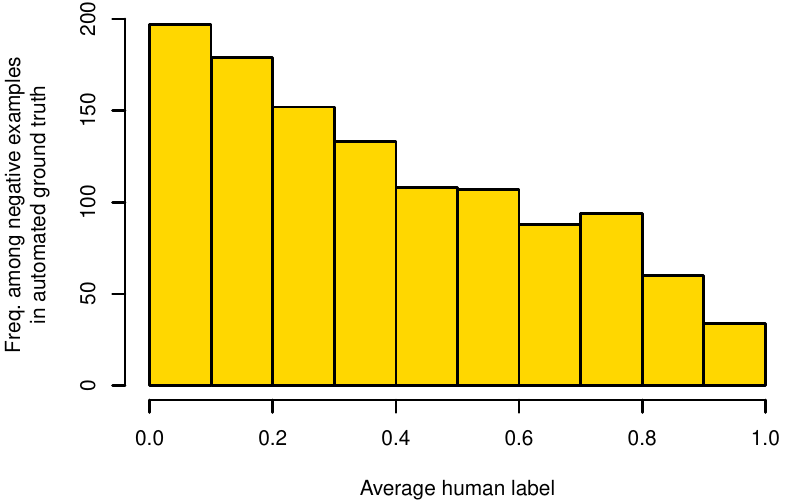}
	\caption{Histogram of average human labels for examples labeled as negative by the automatically obtained ground truth,
	highlighting the prevalence of false negatives in the latter.}
 \label{fig:false_negatives_distribution}
\end{figure}

\xhdr{False negatives in the automated ground truth}
Now that we have human-labeled data, we can quantify the prevalence of false negatives in the automatically constructed ground truth.
For this purpose, consider \Figref{fig:false_negatives_distribution}, which shows a histogram of the average human labels for the candidates that were labeled as negative according to the automatic ground truth.
Here, `average human label' refers to the average of the binary labels obtained from the ten human raters for each candidate.
We see that a large fraction of the examples labeled as
negative according to the automated ground truth are in fact positive examples according to the more reliable human ground truth.

\subsection{Evaluation on Wikispeedia}
\label{sec:Evaluation Wikispeedia}

\noindent
To conclude the evaluation, we present results on the second data set of navigation paths, collected via Wikispeedia.
Recall from \Secref{sec:the_wiki_game} that, while The Wiki Game has the advantage of more data, it also has a slight drawback:
it does not use a static Wikipedia snapshot but rather fetches articles from live Wikipedia on the fly and caches them for some time \cite{clemesha2014}, which means that we do not know the exact version of the article the user saw at game time; the versions used in different games may be different from each other and from the reference snapshot $\mathcal{W}_T$.
Hence, our evaluation on The Wiki Game could not account for what a source article $s$ looked like at game time.
Instead, we allowed for suggestion all links $(s,t)$ not present in the reference snapshot $\mathcal{W}_T$, regardless of whether they existed during the respective game,
and our automated evaluation counted a suggestion $(s,t)$ as positive if the link was present for a substantial fraction of the entire lifetime of $s$.
We call this the \emph{standard evaluation.}

Wikispeedia, on the contrary, uses a static Wikipedia snapshot $\mathcal{W}$~\cite{schools-wikipedia}, so we know exactly which links existed during the game.
In the notation of \Secref{sec:candidate_selection} and \Figref{fig:pipeline}, the snapshot $\mathcal{W}_T$ is replaced by $\mathcal{W}$, which is identical to the snapshot used in all games.
This in turn allows for a stricter evaluation methodology: By allowing for suggestion only those links that were not present in $\mathcal{W}$, we permit only links that did not exist during the game and that the user could thus not possibly have clicked.
Further, we count a suggestion as positive only if it has been present in the live Wikipedia for a substantial amount of time \emph{after} the date of the static snapshot $\mathcal{W}$.
If a suggested link did not exist during the game, but was added afterwards, this is an even stronger signal that the suggestion is good.
Hence we call this the \emph{stricter evaluation}.%
\footnote{When applying the standard evaluation to Wikispeedia, we use the same reference snapshot $\mathcal{W}_T$ also used for The Wiki Game in order to decide whether a source mentions, or links to, a target. Under the stricter evaluation, the static snapshot $\mathcal{W}$ is used for this purpose.}

Now, if we can show that the stricter evaluation yields similar results to the standard evaluation on Wikispeedia, then we may argue by analogy that the stricter evaluation would likely give similar results on The Wiki Game, too, if such an evaluation were possible on that data set.

The results are displayed in \Figref{fig:prec_at_k_wikispeedia_current} and \Figref{fig:prec_at_k_wikispeedia_schools}.
As for The Wiki Game, we use the link-rate threshold (\Secref{sec:weak_ground_truth}) $\alpha=30\%$ for deciding if a suggestion is positive, and as described in the beginning of \Secref{sec:evaluation}, we consider targets for which our algorithm can make at least $K=5$ suggestions and show the average precision@$k$ for $k=1,\dots,5$.

Our first observation is that the standard\hyp evaluation results are similar for Wikispeedia (\Figref{fig:prec_at_k_wikispeedia_current}) and The Wiki Game (\Figref{fig:prec_at_k}).
In particular, the orderings of methods by performance are identical.
(The results are somewhat less clean for Wikispeedia, due to the smaller amount of data.)
Further, the outcome of the stricter evaluation (\Figref{fig:prec_at_k_wikispeedia_schools}) is similar to that of the standard evaluation (\Figref{fig:prec_at_k_wikispeedia_current}),
the main difference being that SVD ranking performs better under the stricter evaluation.

We therefore have reason to believe that the performance would also remain high on The Wiki Game under the stricter evaluation
if this kind of evaluation were possible on that data set, which corroborates the result that our algorithm finds good new links.

\section{Discussion}
\label{sec:discussion}

\noindent
This paper introduces an effective method for the source prediction problem (\Figref{fig:motivation}(b)), in which a target page $t$ is given, and the task is to find and rank sources $s$ that should link to $t$.
Prior work (\Secref{sec:relatedwork}) has primarily addressed the complementary target prediction problem (\Figref{fig:motivation}(a)), where $s$ is given and $t$ to be found.
We consider source prediction more challenging than target prediction, since in the latter the set of link candidates is immediately given by the phrases contained in the source page $s$, whereas, in the former, every page could potentially be a source, so the set of source candidates must first be retrieved in a candidate selection step.

\xhdr{Computational feasibility}
To illustrate this point, we briefly report on an experiment we had initially planned on doing.
We intended to compare the performance of our method to the link predictions made by Milne and Witten's \cite{milne+witten2008_link} machine\hyp learned target prediction algorithm, but this was computationally infeasible: In order to use this target prediction method in a source prediction setting, we first had to find all articles $s$ mentioning $t$ (this required a full scan of a 44GB Wikipedia dump).
Next, we intended to annotate each source $s$ with outgoing links and then rank $s$ according to the score it gives to $t$.
However, each annotation takes on the order of several seconds \cite{milne2013open}, and nearly every article mentions at least one of the targets we want to evaluate, so we would have had to annotate essentially all of Wikipedia, which would have taken several million seconds, or several thousands of hours.
One reason for the computational complexity of Milne and Witten's algorithm is that they (as well as other target prediction methods \cite{meij2012adding,mihalcea+csomai2007,wu+weld2007}) tend to spend significant effort on mention disambiguation.

On the contrary, in our approach we neither have to scan Wikipedia for articles that mention $t$, nor do we need to do any sophisticated disambiguation or ranking.
We simply use as source candidates all pages seen in our navigation traces, look for mentions only in this small subset of all Wikipedia pages, and rank according to a simple precomputed metric or simple frequency counts.
This is possible because the brunt of the computational effort is done by humans: since they actively seek out pages that are likely to link to the target, these pages tend to already be good source candidates, and issues such as disambiguation are much less critical.

\xhdr{Applications beyond Wikipedia}
Now we address the question if and how our technique could apply beyond the realm of Wikipedia.
We envision two ways forward.

The first idea would be to gamify arbitrary websites. One could imagine a framework, \eg, written in JavaScript, that would wrap the website of interest, recruit players, and ask them to navigate to the targets we are interested in linking to.
This would require adding at least some initial links pointing to $t$ manually, such that $t$ is reachable by navigating.
Furthermore, our method for finding valid anchors for the target, which is currently based on anchor-text\slash target-page pairs mined from Wikipedia (\Secref{sec:candidate_selection}), would need to be adapted to the new domain.
Possibilities would include the use of prevalent phrases from the target's title and content as anchor texts, or, akin to our current method, the use of anchor texts that are already being used in other pages to refer to the target.

The second approach we envision is to use passively rather than actively collected log data for source candidate selection and ranking.
It might be possible to simply use the logs that are kept by webservers anyway.
The added challenge here would be that we do not know what target (if any) a user tried to reach, whereas the target is always given explicitly to the user in the human\hyp computation setup.
However, we believe that reasoning along the following lines might be promising:
if users that ended up in $t$ often went through $s$, then the shortcut from $s$ to $t$ might be promising.
An alternative heuristic might be to collect instances where a user navigates to $s$, issues a keyword query into the website's search box (if it exists), and clicks to $t$ from the search\hyp engine result page.

On Wikipedia, the linking guidelines are explicitly stated \cite{wikipedia:linking}, so links are fairly consistent. Further, each page is typically about a single, well\hyp defined topic.
These are among the reasons why machine\hyp learning methods can infer powerful models for linking to Wikipedia articles.
Websites other than Wikipedia are less likely to have the above properties, so it will be more difficult for statistical models to predict meaningful links.
We expect methods for mining missing links directly from navigational traces to suffer less from this problem, since they do not take the detour through modeling the static structure of the link graph, but instead directly optimize navigability as the objective.

What we find especially promising in this light is a result from \Figref{fig:prec_at_k_ALL}, namely that our method does not crucially rely on any measure of relatedness between pages: ranking our source candidates simply by the frequency with which they occurred in navigational traces for the given target (the yellow curves of \Figref{fig:prec_at_k_ALL}) constitutes a competitive method.
We believe that this makes our approach a strong candidate for the source prediction task on websites other than Wikipedia, where a notion of relatedness between pages might be much harder to obtain.

\section{Related work}
\label{sec:relatedwork}

\noindent
We find missing links by observing humans navigating a network
during
a human-computation game.
There has been related work on several aspects of our approach:
the link prediction problem, human network\hyp navigation behavior, and games with a purpose.

\xhdr{Link prediction}
The link prediction problem in networks comes in many flavors and variants.
Unsupervised methods for link prediction in social networks were extensively evaluated by Liben\hyp Nowell and Kleinberg~\cite{liben2007link}, who found the Adamic--Adar measure~\cite{adamic03} to perform best.
More recently approaches based on network community detection~\cite{clauset08hierarchical,henderson09link,kim11completion} and random walks~\cite{backstrom11srw} were considered for predicting missing links. Supervised link prediction~\cite{lichtenwalter2010new} was also studied by the relational\hyp learning community~\cite{popescul03linkpred,taskar03linkpred}, but scalability remains a challenge with these approaches.

While the above works focused mostly on the identification of missing links in social networks, there is also a rich line of work on the identification of missing links among Wikipedia articles~\cite{adafre+derijke2005,noraset2014adding,west-et-al2009a,wu+weld2007} and on linking existing webpages to Wikipedia~\cite{meij2012adding,mihalcea+csomai2007,milne+witten2008_link}. Generally these approaches focus on building models of Wikipedia's graph structure, while also performing keyword extraction and word\hyp sense disambiguation.

\xhdr{Human navigation in networks} This line of research is rooted in Milgram's seminal small-world experiment \cite{milgram67smallworld}, which asked participants to forward a letter to a friend such that it could finally reach a predestined target person. The game through which our data was collected is similar to this task in that a target must be reached in the absence of prior information about the underlying network. It is different in that our setup has the same user staying in control from the start all the way to the target, whereas, in the small-world experiment, every step is executed by a new, autonomous participant. Kleinberg \cite{kleinberg2000} investigated the algorithmic aspects of the small-world problem, showing that efficient search is only possible when the probability of long-range links decays as a power law with a specific exponent. 

Much research has followed in Kleinberg's wake, so we focus on the most directly related projects:
data sets such as ours were previously analyzed by West and Leskovec, who characterize human strategies in successful navigation tasks \cite{west+leskovec2012www} and train machine\hyp{}learning models capable of navigating automatically \cite{west+leskovec2012icwsm}, and by Helic \etal~\cite{helic2013models} and Trattner \etal\ \cite{trattner2012exploring}, who explore heuristic navigation algorithms based on hierarchical knowledge representations.

A related line of work pertains to the analysis of so-called `click trails'.
Research here primarily studies the click paths on which users embark starting from search-engine result pages. In early fundamental work, Chi \etal\ \cite{chi01scent} coin the notion of `information scent', operationalized by Olston and Chi \cite{olston03stenttrails} in a system for supporting users by combining query- and click\hyp{}based navigation strategies.
White and Huang \cite{white10scenic} establish that click trails add value to the information contained in the ultimate target page, and Teevan \etal\ \cite{teevan04teleport} show that users frequently prefer click\hyp{}based navigation to querying. Downey \etal\ \cite{downey2008understanding} investigate the benefits of navigating versus querying further, finding that navigating is particularly useful when the information need is rare.
Work by White and Singla~\cite{white2011finding} is relevant in that it explores how different trail topologies (such as stars, trees, and linear chains) are observed in different search scenarios (informational versus navigational). 
Click trails have also been used to predict whether users will give up in information network navigation~\cite{scaria2014last}, to compute the semantic relatedness of concepts~\cite{singer2013computing,west-et-al2009}, and to identify relevant websites from user activity \cite{bilenko08trails}.
Our work continues this line of work and attempts to use navigational trails as a rich source of data for detecting missing links in networks.

\xhdr{Games with a purpose}
`Games with a purpose' \cite{vonahn+dabbish2008} were popularized by von Ahn and colleagues, a seminal early example being the \textit{ESP Game} for labeling images \cite{vonahn+dabbish2004}.
Wikispeedia was originally designed as a game with a purpose for computing the semantic relatedness between concepts \cite{west-et-al2009}.
Further relevant work was done by Ageev \etal\ \cite{ageev2011find}, who developed a human\hyp computation game for collecting data in which users are asked to find the answers to as many factual questions (\eg, `What is the highest peak in the Western Hemisphere?') as possible within a given amount of time, using Web\hyp search queries that may optionally be followed by click\hyp based navigation. As in our navigation data sets, the goal is explicitly known here, but not in the form of a specific target page but rather in the form of a specific answer string.

\section{Conclusion}
\label{sec:conclusions}

\noindent
In this paper we studied the problem of identifying missing links in Wikipedia. We built on the fact that the ultimate purpose of Wikipedia links is to aid navigation. Our method harnesses human navigation traces and finds missing links that would immediately enhance Wikipedia's navigability. We analyze click trails to identify a set of candidates for missing links and then rank these candidates. We experimented on both automatically labeled ground truth as well as ground truth obtained from human annotators. Overall, we obtained performance of the quality that would make our system useful in practice.

There are many interesting avenues for future work. For instance, extending the method to passively collected Web\hyp browsing logs would be a natural next step, and it would also be worthwhile to think about gamifying general websites beyond Wikipedia.

In summary, our paper makes contributions to the rich line of work on detecting missing links on websites. We hope that future work will draw on our insights to build more user\hyp friendly websites and make the Web more navigable as a whole.


{\small
\xhdr{Acknowledgments}
This research has been supported in part by NSF
IIS-1016909,              
IIS-1149837,       
IIS-1159679,              
CNS-1010921,              
NIH R01GM107340,
Boeing,                    %
Facebook,
Volkswagen,                 
Yahoo, and
SDSI.
Robert West acknowledges support by a Stanford Graduate Fellowship and a Facebook Graduate Fellowship, and thanks the Wikimedia Foundation for sharing their office space.
Finally, we would like to thank Alex Clemesha for sharing data from The Wiki Game.
}

\appendix
\section{Human-rater instructions on\\Amazon Mechanical Turk}
\label{app:instructions}
\noindent
The following description of the evaluation task was given to human raters on Amazon Mechanical Turk:

{\em
``Here's the deal! Our good friend Wikipedia is having self-doubts and wants you to help improve its links.

You are given a Wikipedia article (referred to as the target) and a list of other Wikipedia articles (referred to as source articles). You have to tell Wikipedia if the source article should contain a link to the target. And of course, if you are unsure of what the source or target article means, you can always click on the article name to open it in a new tab.

But remember that Wikipedia is a sensitive fellow and will be mad if you don't play by the rules: There should be a link from the source to the target if and only if
(1)~the target article has some relevant information about the source article and could help readers understand the source
more fully, or
(2)~the target article describes a proper name which is likely to be unfamiliar to readers.''

}
\end{document}